\documentclass[mathleft]{an}
\usepackage{graphicx}
\usepackage{times}
\def \f#1{\ensuremath{\vec{#1}}}
\def \n  {\vec{\nabla}\!}
\def \p  {\partial}
\def \t  {\times}

\begin{document}

\Pagespan{1}{}
\Yearpublication{2007}%
\Yearsubmission{2007}%
\Month{}%
\Volume{}%
\Issue{}%

\title{Energy distribution in nonaxisymmetric magnetic Taylor-Couette flow}

\author{M. Gellert\inst{1}\fnmsep\thanks{{Corresponding author: mgellert@aip.de}} \and G. R\"udiger\inst{1} \and A. Fournier\inst{2}\newline}
\titlerunning{Energy distribution in nonaxisymmetric magnetic Taylor-Couette flow}
\authorrunning{M. Gellert, G. R\"udiger \& A. Fournier}
\institute{Astrophysikalisches Institut Potsdam, An der Sternwarte 16, D-14482 Potsdam, Germany 
           \and
           Laboratoire de G\'eophysique Interne et Tectonophysique, Universit\'e Joseph-Fourier, CNRS, BP 53, 38041 Grenoble cedex 9, France}

\received{2007}
\accepted{2007}
\publonline{2007}


\abstract{%
         Azimuthal magnetorotational instability is a mechanism that generates nonaxisymmetric field pattern. Nonlinear 
         simulations in an infinite Taylor-Couette system with current-free external field show, that not only the linearly 
         unstable mode $m=1$ appears, but also an inverse cascade transporting energy into the axisymmetric field is possible. 
         By varying the Reynolds number of the flow and the Hartmann number for the magnetic field, we find that the ratio between 
         axisymmetric ($m=0$) and dominating nonaxisymmetric mode ($m=1$) can be nearly free chosen. On the surface of the outer 
         cylinder this mode distribution appears similarly, but with weaker axisymmetric fields. We do not find significant differences 
         in the case that a constant current within the flow is added.
         }

\maketitle

\section{Introduction}

MHD-instabilities in radiation zones of differentially rotating stars influence the rotation profile of the star due to angular momentum 
transport (Denissenkov \& Pinsonneault 2007; Yoon, Langer \& Norman 2006). Tayler instability (Tayler 1973) or helical 
magnetorotational instability (Hollerbach \& R\"udiger 2005) are examples for such instabilities. According to Spruit (1999) the Tayler 
instability could possibly support a dynamo. In their model 
the instability should create a significant poloidal field that due to differential rotation is wound up into a toroidal one and thus close 
the loop for a self-sustained dynamo (Braithwaite \& Spruit 2004; Braithwaite 2006). While this idea sounds promising, there are doubts about its 
general functionality. Simulations in very similar geometry with $z$-dependent differential rotation did not show a dynamo (Gellert, R\"udiger \& Elstner 2007). 
Zahn, Brun \& Mathis (2007) claim that such a process is not possible in this simple way. The authors argue that the nonaxisymmetric modes wound up by differential 
rotation can not contribute to the axisymmetric field, because the azimuthal wave\-number is always $m\neq0$. As we show in the following, 
azimuthal magnetorotational instability (AMRI) creates also an axisymmetric component in the nonlinear regime. For a scenario based on AMRI as instability 
mechanism that might overcome the arguments of the incompatible wavenumber and would make field regeneration at least possible. It does not necessarily mean, 
however, that it supports a dynamo.

This work consists of two aspects. First we describe the development of nonaxisymmetric modes in the nonlinear regime of AMRI-unstable Taylor-Cou\-ette 
flow. Especially the observed inverse cascade transporting energy into the axisymmetric field and its dependence on the shear is analyzed. As second 
aspect we point out that weak currents within the flow do not basically influence the instability. Field structure and energy distribution remain almost 
unchanged.

\section{Equations and numerical treatment}

We use the hydrodynamic Fourier spectral element code described by Fournier et al. (2005) extended by its magnetic field.
With this approach we solve the 3D MHD equations 
\begin{eqnarray}\label{eq_ns}
\lefteqn{ \p_t \f{u}  + (\f{u}\cdot\n)\f{u} = -\n p + \n^{\,2} \f{u} + \frac{\mathrm{Ha}^2}{\mathrm{Pm}} (\mathrm{rot}\f{B}) \t \f{B},} \\
 \lefteqn{ \p_t \f{B}  = \frac{1}{\mathrm{Pm}} \n^{\,2} \f{B} + \mathrm{rot}(\f{u}\t\f{B}),}\\
 \lefteqn{\mathrm{div}\f{u}  = 0, \qquad \mathrm{div}\f{B} =0} 
\end{eqnarray}
for an incompressible medium in cylindrical coordinates \\ $(R, \phi, z)$. Free parameters are the Hartmann number 
\begin{equation}\label{eq_ha}
 \mathrm{Ha} = B^0_\mathrm{in}\sqrt{\frac{R_\mathrm{in} D}{\mu_0\rho\nu\eta}}
\end{equation}
and the magnetic Prandtl number $\mathrm{Pm} = \nu/\eta$. Here $\nu$ is the viscosity of the fluid and $\eta$ its magnetic diffusivity.  
The Reynolds number is defined as $\mathrm{Re}=\Omega_\mathrm{in} R_\mathrm{in} D / \nu$ with  $D=R_\mathrm{out}-R_\mathrm{in}$ (unit of length) and the
angular velocity of the inner cylinder $\Omega_\mathrm{in}$. Unit of velocity is $\nu/D$ and unit of time the viscous time $D^2/\nu$.

The solution is expanded in $M$ Fourier modes in the azimuthal direction. This gives rise to a collection of meridional problems, each of which is 
solved using a Legendre spectral element method (see e.g. Deville, Fischer \& Mund 2002).  Either $M=8$ or $M=16$ Fourier modes are used, three elements in 
radial and eighteen elements in axial direction. The polynomial order is varied between $N=8$ and $N=12$. With a semi-implicit approach consisting 
of second-order backward differentiation formula and third order Adams-Bashforth for the nonlinear forcing terms time stepping is done with 
second-order accuracy (Fournier et al. 2004).

At the inner and outer wall perfect conducting boundary conditions are applied. In axial direction we use periodic boundary conditions to avoid 
perturbing effects from solid end caps. The periodicity in $z$ is set to $\Gamma=6D$.

The initial flow profile is the typical Couette profile
\begin{equation}\label{eq_couette}
 \Omega(R) = a + \frac{b}{R^2}
\end{equation}
with
\begin{equation}
a = \frac{\hat{\mu}_\Omega - \hat{\eta}^2}{1-\hat{\eta}^2}\Omega_\mathrm{in}, \qquad b = \frac{1-\hat{\mu}_\Omega}{1-\hat{\eta}^2}R^2\Omega_\mathrm{in},
\end{equation}
$\hat{\eta}=R_\mathrm{out}/R_\mathrm{in}=0.5$ and $\hat{\mu}_\Omega = \Omega_\mathrm{out}/\Omega_\mathrm{in} = 0.5$. This means the outer cylinder 
rotates with half of the angular velocity of the inner cylinder in the same direction. After the Rayleigh stability criterion $\p_R (R^2\Omega)^2>0$ 
this configuration is hydrodynamically stable. In addition a toroidal external field $\f{B_\mathrm{ext}}=(0,B^0,0)$ is applied. The second component 
$B^0$ has a radial profile similar to the flow:
\begin{equation}\label{bext_prof}
 B^0(R) = a_B R + \frac{b_B}{R}.
\end{equation}
The external field than can be characterized by the number
\begin{equation}\label{mub_relation}
 \hat{\mu}_B = \frac{B^0_\mathrm{out}}{B^0_\mathrm{in}} = \frac{a_B R_{\mathrm{out}} + b_B R_{\mathrm{out}}^{-1}}{a_B R_{\mathrm{in}} + b_B R_{\mathrm{in}}^{-1}}.
\end{equation}
The current-free field is given by $\hat{\mu}_B=0.5$. Values below lead to a field connected with constant negative current $j_z$ 
within the flow and larger values to a positive current. To trigger the instability the initial magnetic field consists of small random perturbations 
(with an amplitude of $10^{-6}B^0_\mathrm{in}$).

\section{Results}
\subsection{Linear instability}

\begin{figure}
\includegraphics[width=0.45\textwidth]{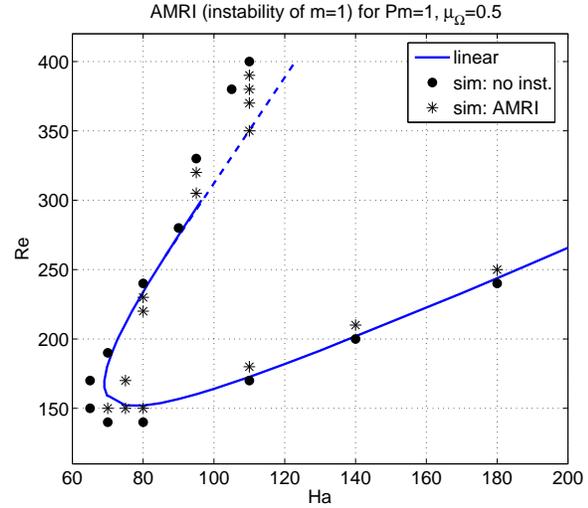}
\caption{Stability diagram for $\mathrm{Pm}=1, \hat{\mu}_\Omega=0.5$ and $\hat{\mu}_B=0.5$ (AMRI, current-free external field). The solid line results from 
         linear stability analysis, dots and stars mark results from nonlinear simulations. The dashed line is an assumed continuation of the solid line, 
         there were no linear data available.}
\label{fig_linstab}
\end{figure}

From linear stability theory is known that only the mode $m=1$ is unstable (R\"udiger et al. 2007a), which is reflected in the nonlinear simulations, too. 
Only $m=1$ starts to grow in the beginning of a simulation where one is in a quasi-linear regime. The stability boundaries found with the nonlinear 
code compared to linear results are shown in Fig. \ref{fig_linstab} for $\mathrm{Pm}=1, \hat{\mu}_\Omega=0.5$ and $\hat{\mu}_B=0.5$. Equal to the 
results in R\"udiger et al. (2007b) we find not only an onset of the instability if the shear (given by the Reynolds number) exceeds a certain threshold for a 
given magnetic field configuration (constant $\mathrm{Ha}$), but also an upper boundary. Strong shear prevents the instability - from this point of view an 
instability of the magnetic field.  Both, the linear results and nonlinear simulations, agree very well on the lower boundary. 
At the upper boundary for $\mathrm{Re}>300$ the simulations exhibit a shift of the stability boundary to slightly higher values of $\mathrm{Re}$. But the 
qualitative picture is the same.

Following a line with constant $\mathrm{Re}$, one can argue that weak but not to small magnetic fields lead to an instability of the flow. For higher 
$\mathrm{Re}$ also the magnetic field has to exceed a higher threshold to sustain the nonaxisymmetric mode generation. Insofar both descriptions, unstable 
flow due to a magnetic field and magnetic field instability due to shear, are equivalent.

\subsection{Nonlinear simulations}
\subsubsection{Current-free external field}

\begin{figure}
\includegraphics[width=0.45\textwidth]{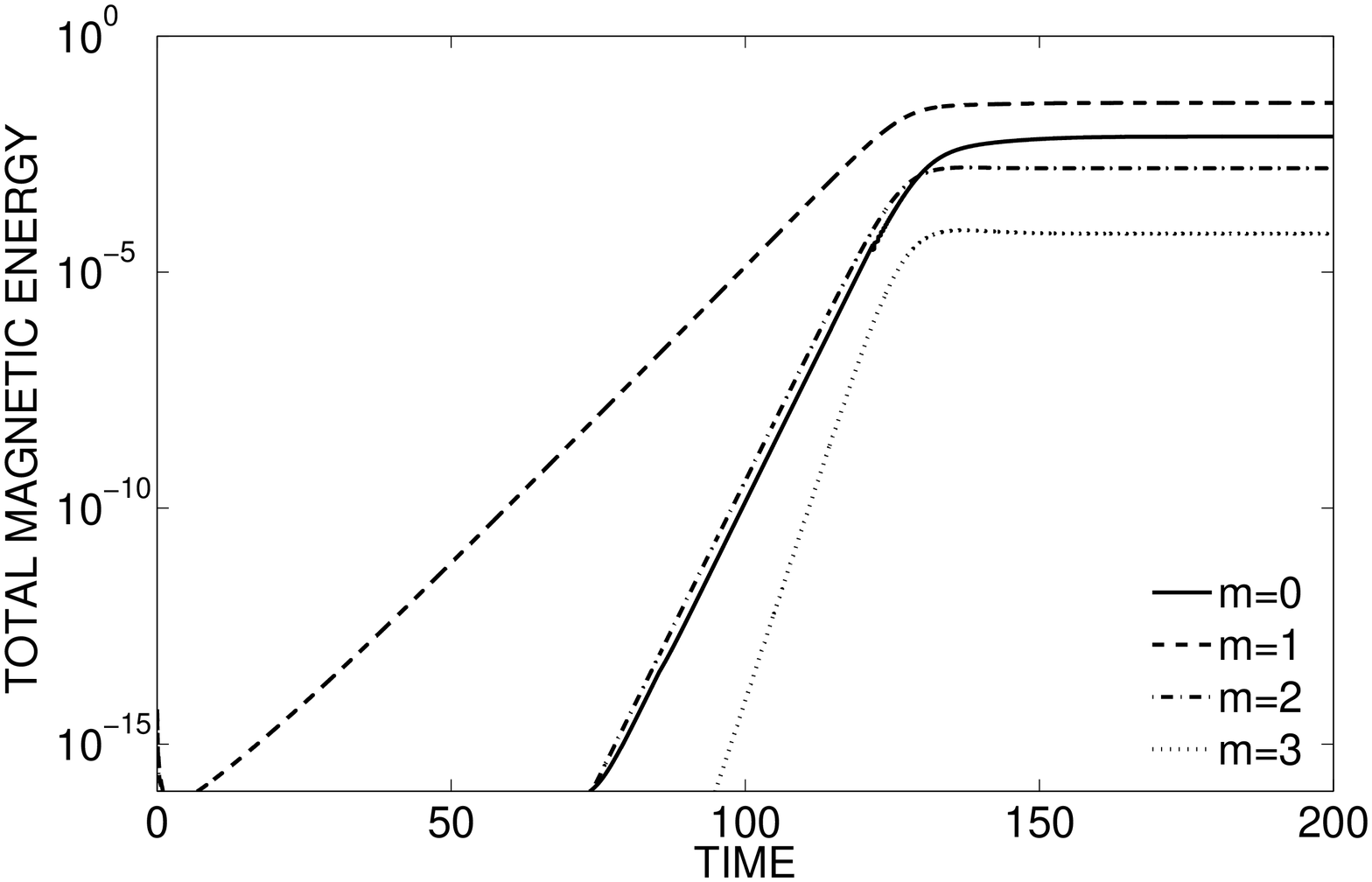}
\caption{Time series of the total magnetic energy for Fourier modes $0\leq m \leq 3$ for $\mathrm{Re}=200$. Linearly unstable is only $m=1$. Time is scaled 
         in periods of rotation.}
\label{fig_ts_mag}
\end{figure}

\begin{figure}
\begin{minipage}[c]{0.19\textwidth}
 \includegraphics[width=0.95\textwidth]{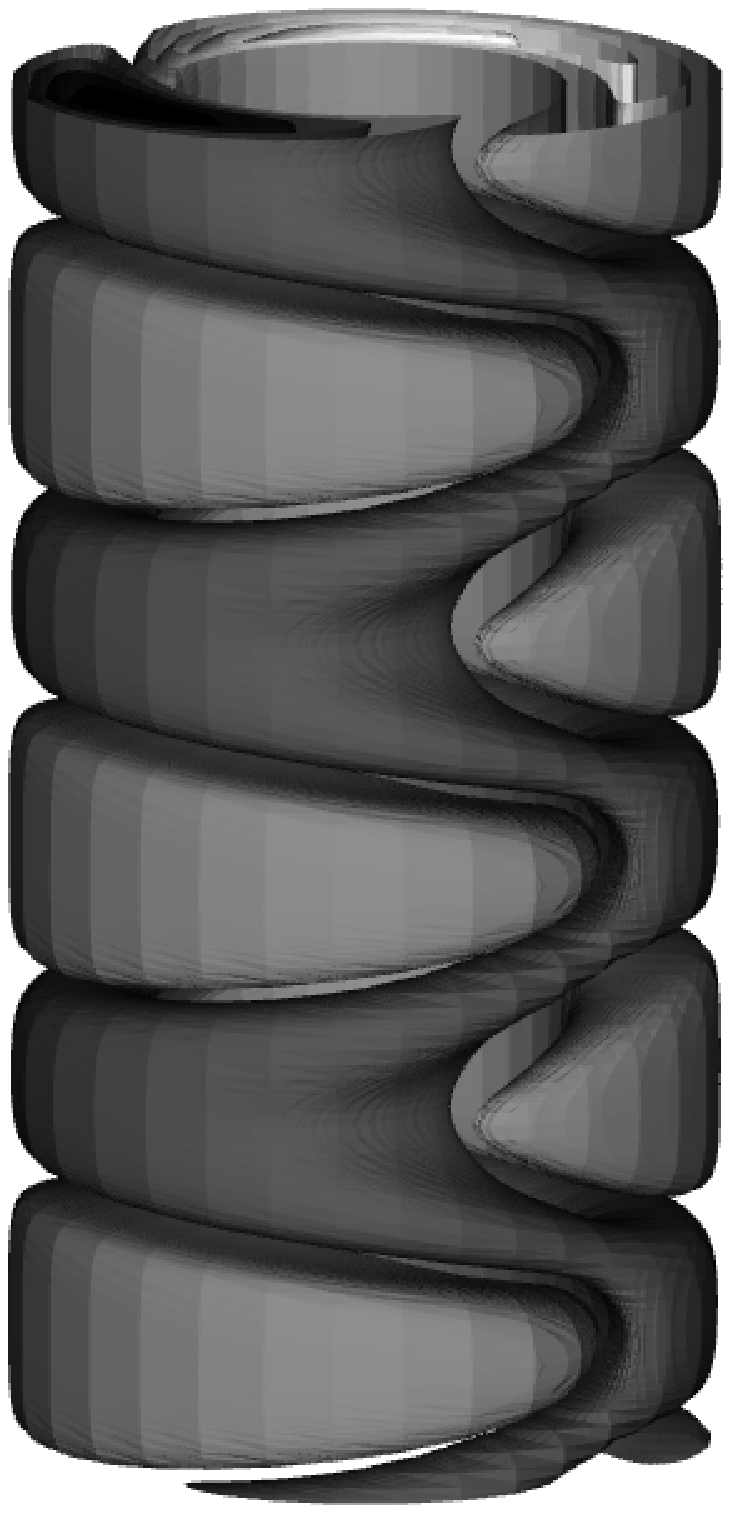}
\end{minipage}
\begin{minipage}[c]{0.29\textwidth}
 \includegraphics[width=0.95\textwidth]{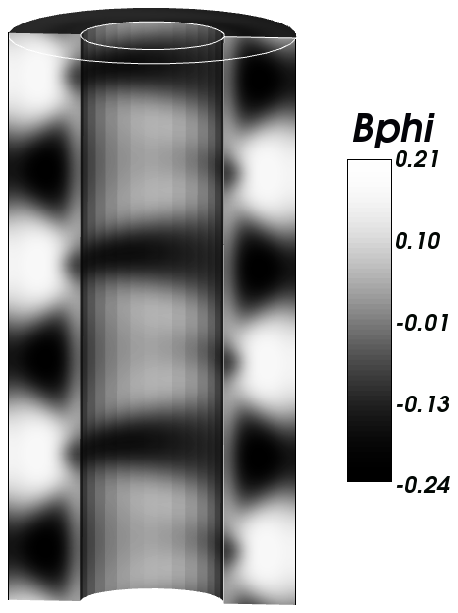}
\end{minipage}\\[1mm]
\caption{On the left an isosurface plot of $B_R$ ($\pm0.7|B_R|_{\mathrm{max}} = 0.09 B^0_\mathrm{in}$) for $\mathrm{Re}=200, \mathrm{Ha}=110$ at $t=300$ is 
         shown, on the right the magnetic field component $B_{\phi}$ in the $R$-$z$-plane also in units of $B^0_\mathrm{in}$.}
\label{fig_cont}
\end{figure}

\begin{figure}
\includegraphics[width=0.23\textwidth]{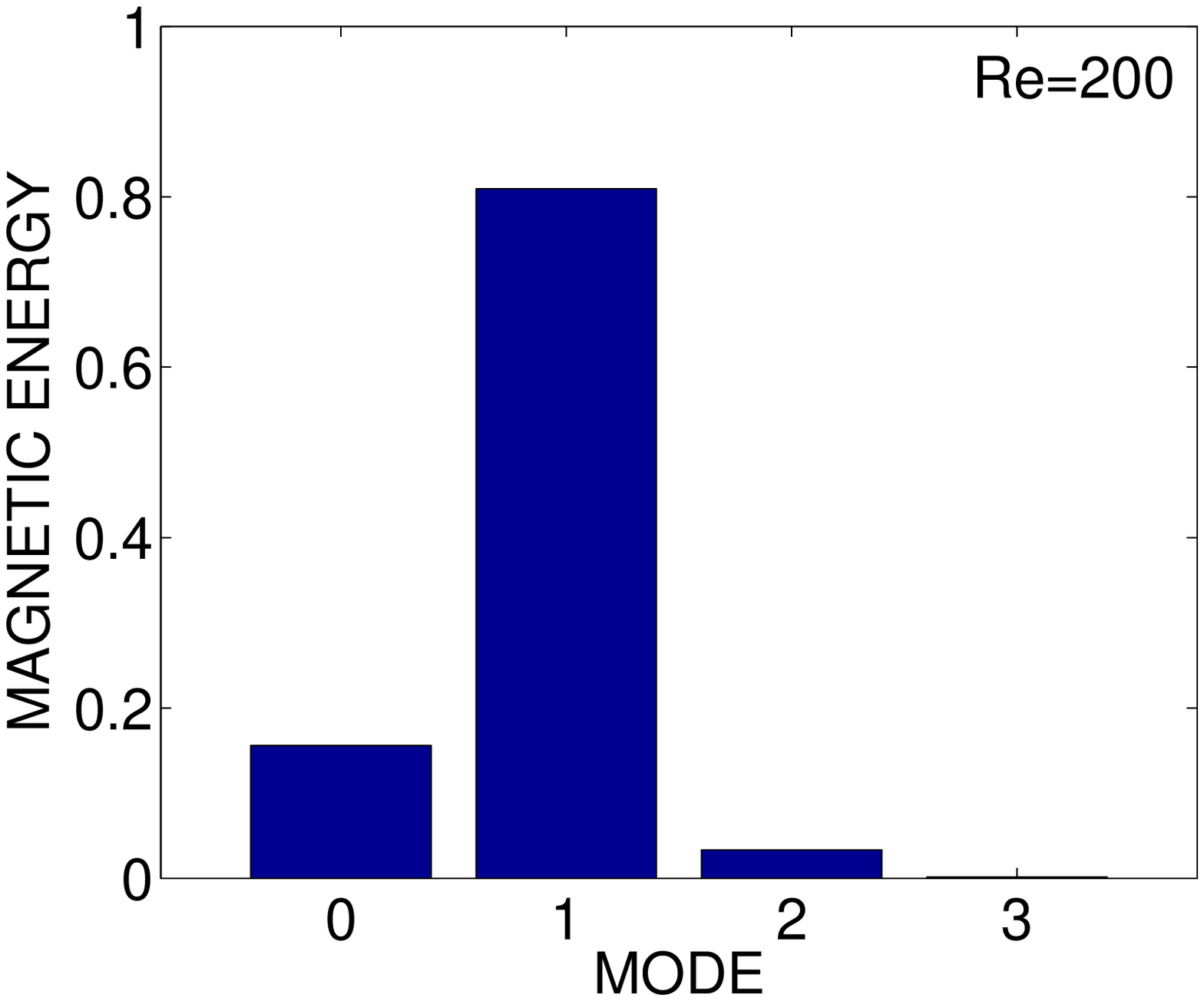}
\includegraphics[width=0.23\textwidth]{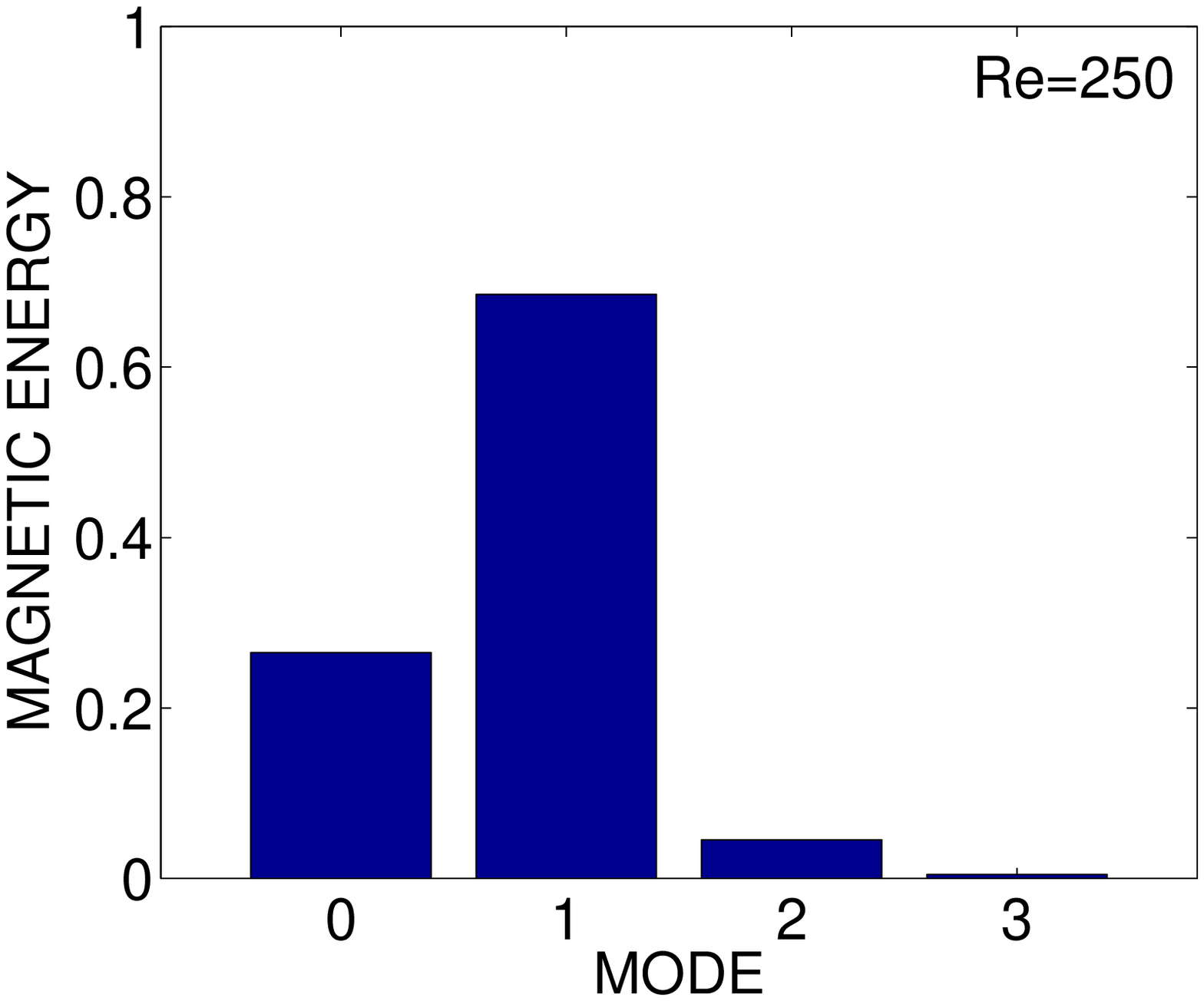}\\
\includegraphics[width=0.23\textwidth]{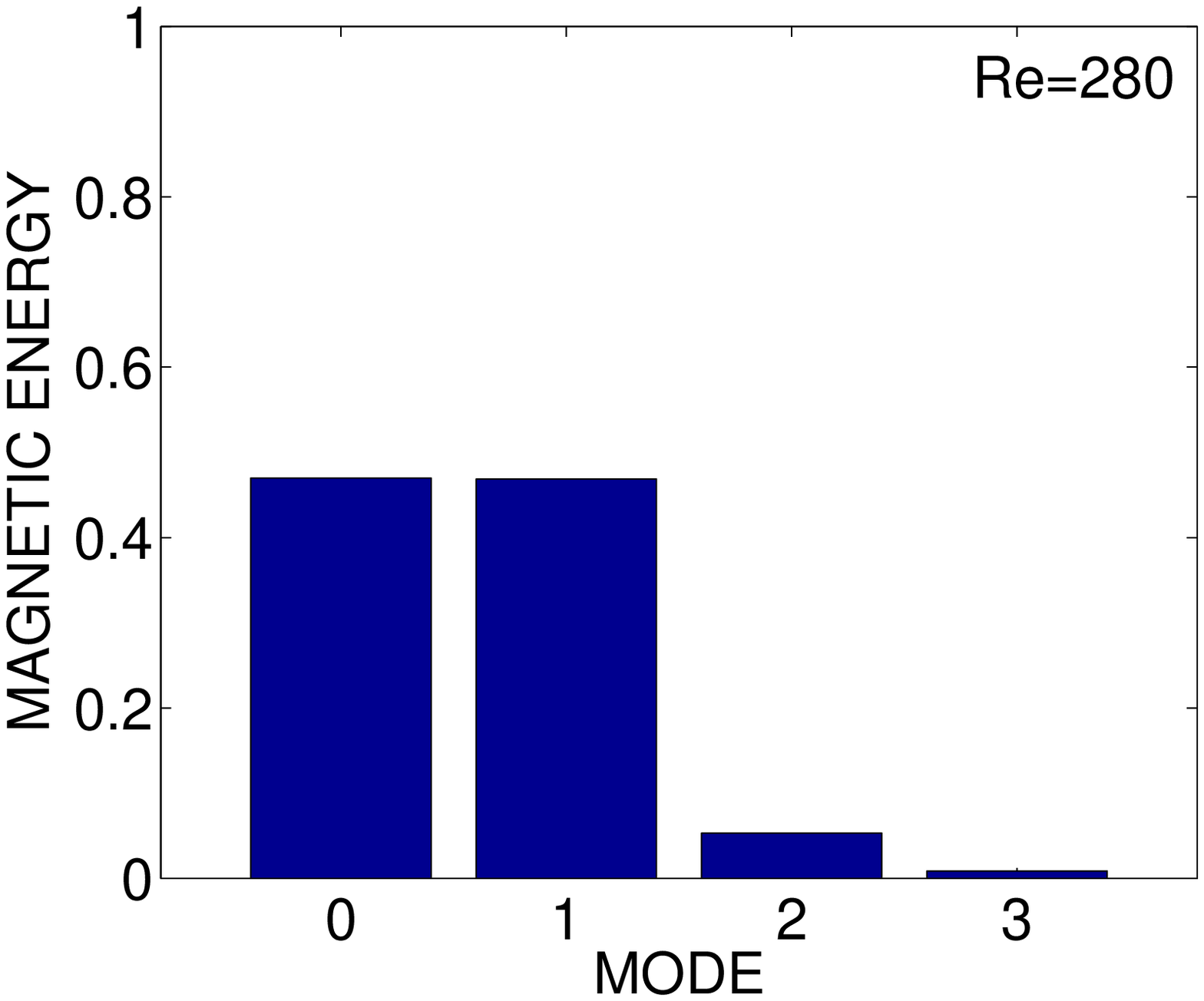}
\includegraphics[width=0.23\textwidth]{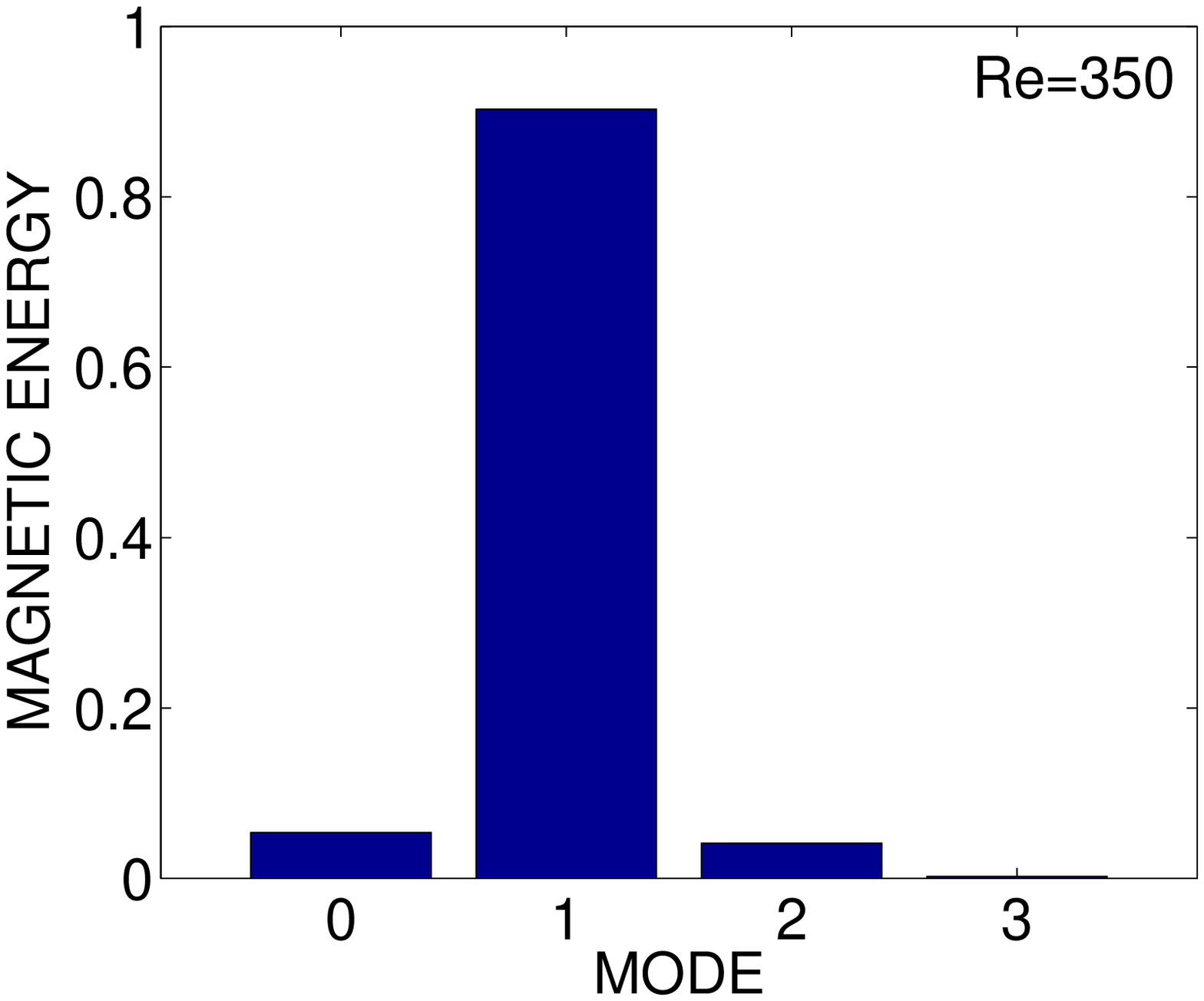}
\caption{Total magnetic energy distribution in Fourier modes $m$ with growing shear for $\mathrm{Ha}=110$. The instability is observable between
         $\mathrm{Re}=175$ and $\mathrm{Re}=390$.}
\label{fig_shear}
\end{figure}

After the onset of the instability, first exclusively the $m=1$ mode grows consistent with the linear theory. If the $m=1$ mode reaches roughly
30\% the amplitude of the final state, also both neighboring modes $m=0,2$ start to grow. Slightly later successively higher and higher modes follow. 
The energy drops by a factor of ten for each $m\rightarrow m+1$. Modes $m>3$ carry less than $10^{-3}$ of the total energy and are not plotted in 
Fig. \ref{fig_shear}. A time series of the total magnetic energy 
\begin{equation}
E(m) = \frac{1}{2V} \int_V \f{B}^2(m)\; \mathrm{d}V
\end{equation}
\label{emag}
for each of the significant modes is shown in Fig. \ref{fig_ts_mag}, the flow and magnetic field pattern is shown in Fig. \ref{fig_cont}. Compared to the strength 
of the external field, the magnetic field due to the AMRI is weaker. The maximum size is around 0.3 $B^0_{\mathrm in}$ and the energy below 10\%. The 
saturated state shows no regular time-dependence, the field pattern rotates with a small drift relative to the system rotation (as already pointed out 
by the linear analysis).

Interesting is the fact that in the nonlinear simulations the modes $m>1$ appear only very weak. The energy of $m=5$ is already by a factor of $10^5$ less 
than the energy of the mode $m=1$.  On the other hand also a significant large mode $m=0$ develops. This means energy transport happens not only to higher 
modes or smaller scales but also in the opposite direction \footnote{Regarding a dynamo based on the instability, Zahn et al. (2006) point out that the generation of an axisymmetric field from a 
nonaxisymmetric one is difficult. If the instability itself also creates an axisymmetric mode the generation of an appropriate poloidal axisymmetric field might
be easier to achieve.}. 

\begin{figure}
\includegraphics[width=0.48\textwidth]{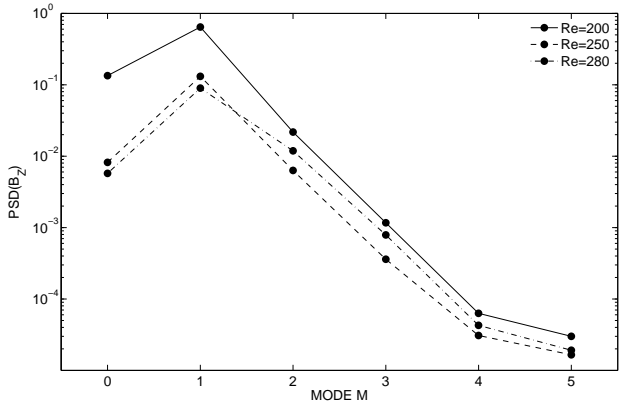}
\caption{Power spectral density for $B_z$ on the cylinder surface for several $m$ normalized by the total power on the surface. The
         distribution is similar to the volume-averaged values, except the $m=0$ mode occurs to be smaller in $B_z$.}
\label{fig_surf}
\end{figure}

By varying the shear strength, the amount of energy in the produced axisymmetric mode can be nearly free chosen. Near the stability boundary 
(e.g. $\mathrm{Re}=175$ and $\mathrm{Re}=390$ for $\mathrm{Ha}=110$, see Fig. \ref{fig_linstab}) the nonaxisymmetric mode $m=1$ dominates the energy spectrum, 
in between at $\mathrm{Re}=280$ $m=0$ becomes of the same size. This seems to be a general behavior and is not only the case for $\hat{\mu}_\Omega=0.5$. 
Also for steeper rotation laws, e.g. the quasi-keplerian ($\hat{\mu}_\Omega=0.3535$), this kind of energy transport into the axisymmetric mode is observable 
in the simulations. The amount of energy in the axisymmetric mode is always of the order of $50$\% of the whole energy or less. This agrees with 
magnetic fields observed in Ap stars.  According to this observations slow rotating Ap stars with strong fields often possess a nonaxisymmetic field. And the 
axi- and nonaxisymmetric modes are of the same size (Landstreet \& Mathys 2000; Bychkov, Bychkova \& Madej 2005).
Regarding these observations more interesting than the volume averaged energy is the behavior of the magnetic field that penetrates the surface and
would be visible from outside. This is the poloidal part of the field, which consists due to the boundary conditions only of the $z$-com\-ponent. Looking 
on the poloidal field, the situation is similar to the global field distribution. The dominating mode is yet the $m=1$ mode. Different is the behavior of the 
produced axisymmetric mode. It is nearly not visible on the surface. And opposite to the global behavior it becomes even less with Reynolds numbers in 
the middle of the unstable parameter space, where the energy of the $m=0$ mode is maximal in the volume-averaged quantity. This is illustrated in 
Fig.~\ref{fig_surf}, where the power spectral density $|\tilde{B}_z(m)|^2$ at $R=R_\mathrm{out}$ and $z=1/3\Gamma$ is plotted for different Reynolds numbers. 
The created axisymmetric component contributes mainly to the $B_\phi$-component, the original source of the instability. Due to the perfect conducting boundary
conditions, which are quite different to the vacuum environment for real stars, the surface field is not directly comparable with the observations.

\subsubsection{External field connected with a current}

With the profile of the external field as defined in Eq. (\ref{bext_prof}), an associated current is always constant. It is positive if $\hat{\mu}_B>0.5$ and negative if $\hat{\mu}_B<0.5$. To test the influence of its current on the appearance of the instability and 
the resulting  energy distribution, we alter the external field profile to $\hat{\mu}_B=0.35$ and $\hat{\mu}_B=0.65$. These modifications of the external field 
are small enough not to change the profile qualitatively. The simulations with the modified external fields do not show a basic influence in the energy 
distribution from a current. The ratio between magnetic energy of modes $m=0$ and $m=1$ has values of $E_0/E_1=0.37$ for $\hat{\mu}_B=0.35$, $E_0/E_1=0.39$ 
for $\hat{\mu}_B=0.50$ and $E_0/E_1=0.40$ for $\hat{\mu}_B=0.65$. With increasing $\hat{\mu}_B$ also the energy of the external field increases and is 
responsible for the slightly increasing energy ratio. The instability is not strongly effected by low currents. For larger currents, or larger Hartmann 
numbers, the influence of the Tayler instability grows and the Reynolds number dependence of the onset of the instability is lost. Tayler instability already 
works with low or even without shear.

\begin{figure}
\includegraphics[width=0.15\textwidth]{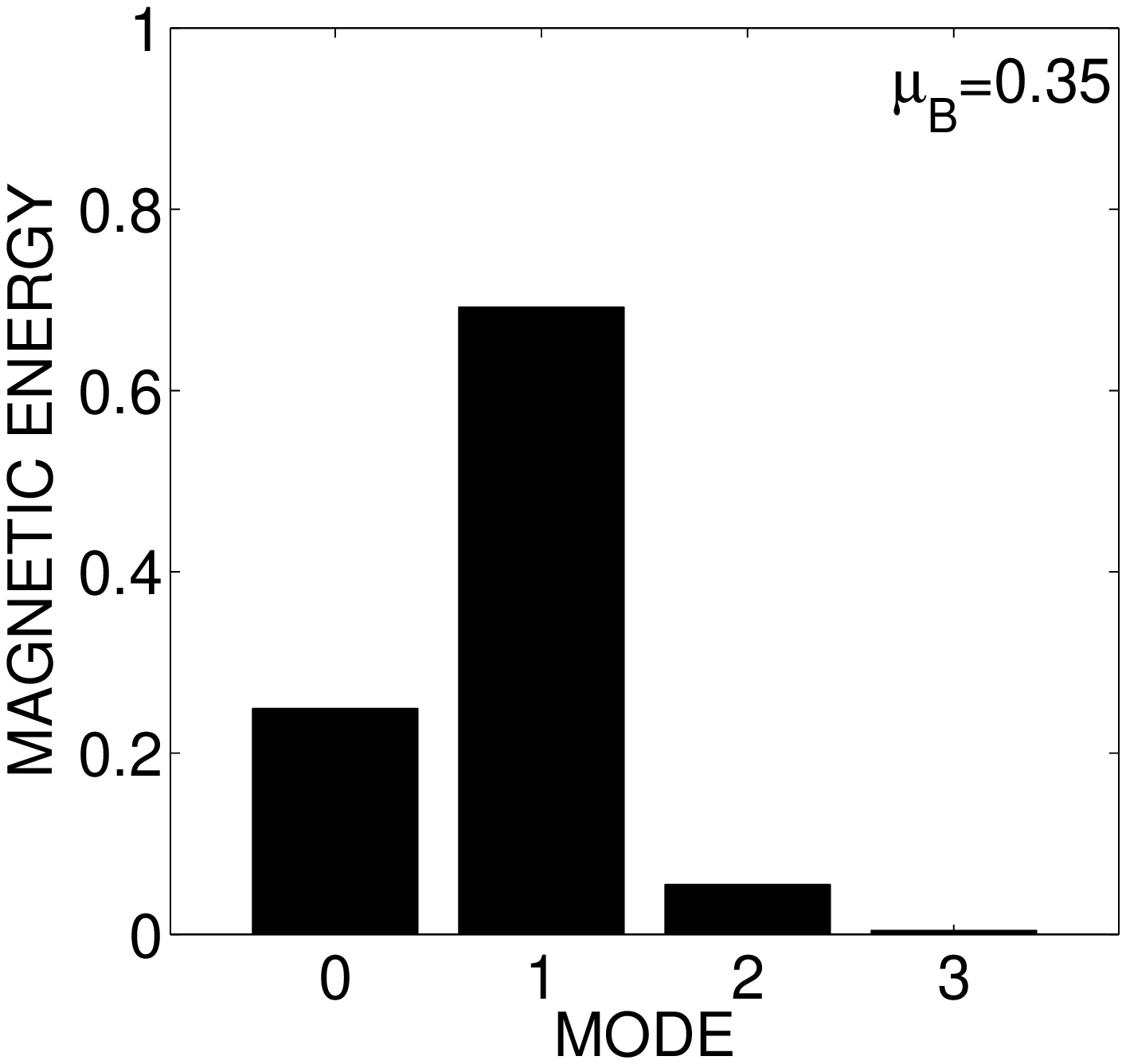}
\includegraphics[width=0.15\textwidth]{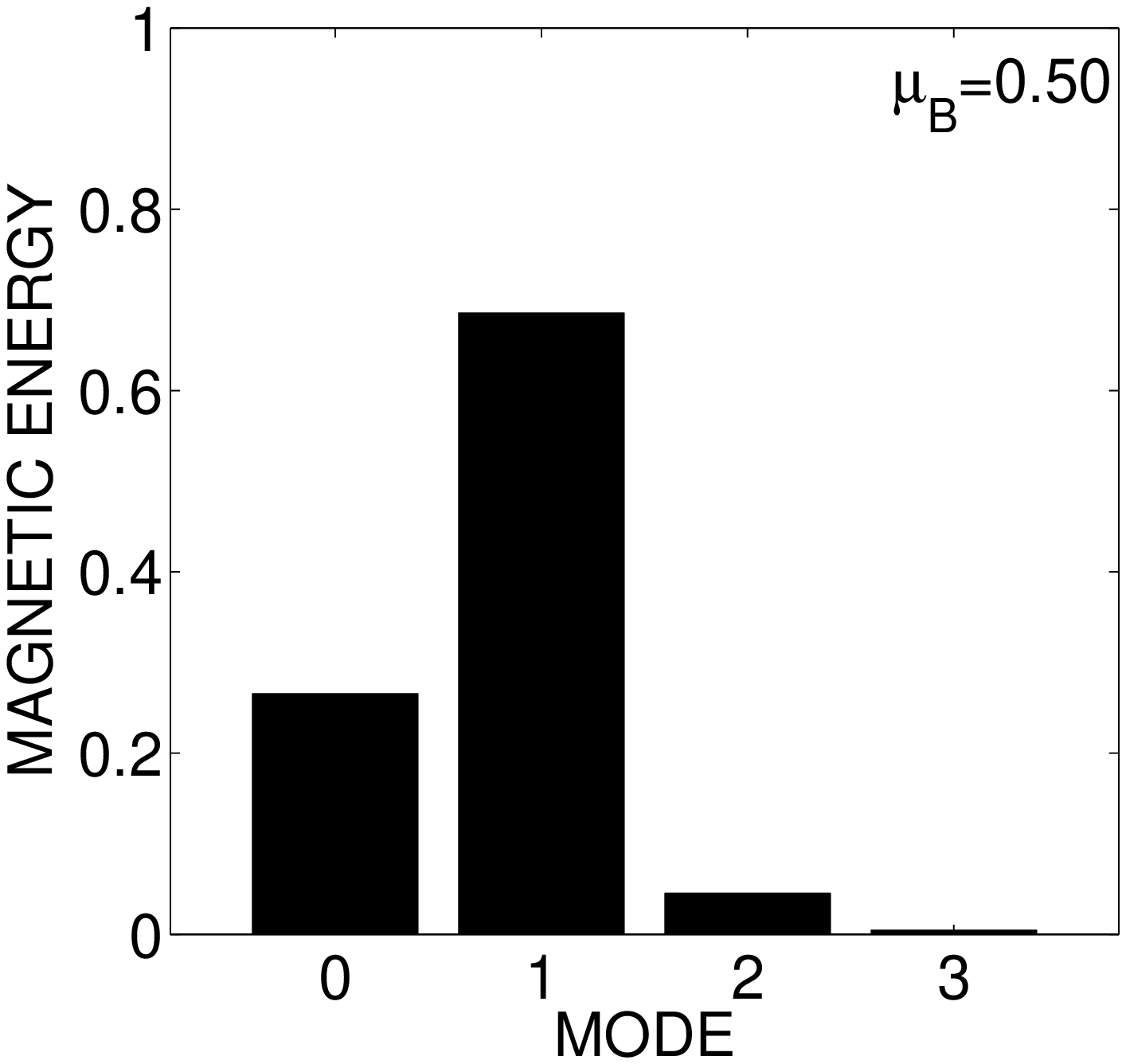}
\includegraphics[width=0.15\textwidth]{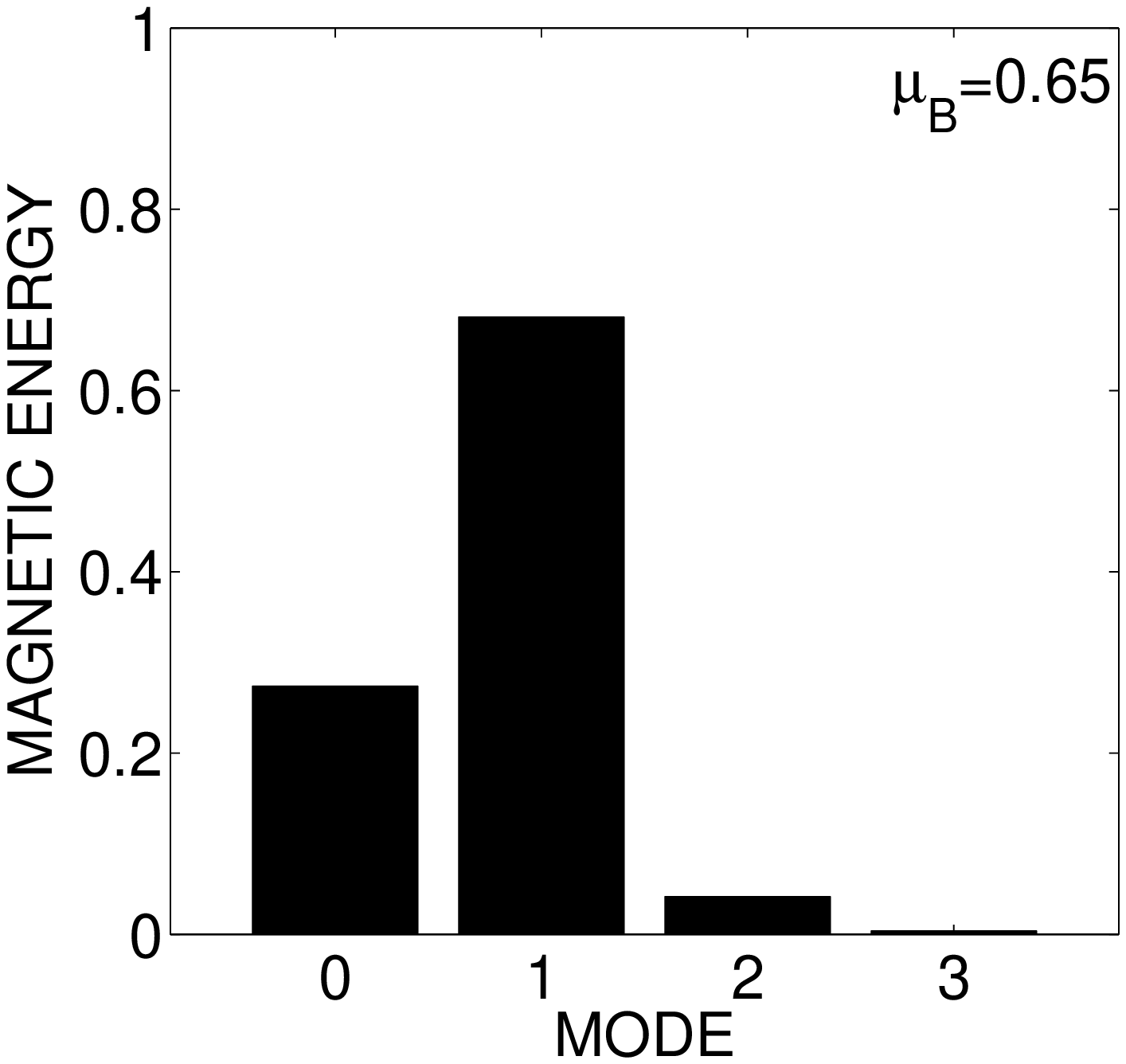}
\caption{Influence of a positive or negative current on the energy distribution ($\hat{\mu}_B=0.35,0.5,0.65$ from left to right) is not visible. The axisymmetric 
         mode carries $E_0=0.249,0.265,0.272$ \% of the total energy, the $m=1$ mode $E_1=0.690,0.685,0.682$ \% for $\hat{\mu}_B=0.35,0.5,0.65$. Histograms are 
         normalized with total magnetic energy.}
\label{fig_current}
\end{figure}

\section{Conclusion}
In an infinite Taylor-Couette system we have shown, that nonlinear simulations of AMRI lead to an unstable nonaxisymmetric
mode $m=1$ of the magnetic field if a certain strength of the external magnetic field is reached. Furthermore the instability is suppressed if the external field
becomes too strong (for given Reynolds number). Equivalent is the point of view of an instability of the magnetic field due to shear. Here also a certain strength
of the shear or value of the Reynolds number has to be reached to trigger the instability of the $m=1$ mode. And if the shear exceeds a certain value, the 
instability is not observable anymore.

The simulations reveal in the saturated state a distribution of magnetic energy over several modes, not only the $m=1$ is excited.
However, the energy transport works in both directions, not only in the small-scale direction to larger $m$, but also to lower $m$. A significant amount of energy is 
shifted inversely into the axisymmetric mode ($m=0$). The shear strength is responsible for the ratio between magnetic energy in axisymmetric and nonaxisymmetric 
mode $m=1$. Near the stability boundaries, the nonaxisymmetric mode carries nearly all energy. In the central region of the unstable parameter space the energy 
ratio between modes $m=0$ and $m=1$ is roughly one. 

Adding a small positive or negative current within the flow does not affect the instability, the field pattern and final energy distribution do not change
compared to the current-free case. A small deviation from the profile $B^0 \propto R^{-1}$ does not visibly influence the occurrence of the instability.


\end{document}